\documentclass[aps,floatfix,showpacs,twocolumn,pre]{revtex4-1}
\usepackage{graphicx} 

\def\g{g}

\def\eref#1{(\ref{#1})}

\begin{document}

\title{Bias- and  bath-mediated pairing of particles  driven through a  quiescent medium}
\author{Carlos~Mej\'{\i}a-Monasterio${}^1$}
\email{carlos.mejia@upm.es}
\author{Gleb Oshanin${}^2$}
\email{oshanin@lptmc.jussieu.fr}
\affiliation{Department of Mathematics and Statistics,
    University of  Helsinki, P.O.  Box 68  FIN-00014 Helsinki, Finland
    \&  Laboratory   of  Physical  Properties,   Department  of  Rural
    Engineering, Technical University of Madrid, Av.  Complutense s/n,
    28040     Madrid,     Spain.${}^1$}
\affiliation{Laboratoire de Physique Th\'eorique de la
    Mati\`ere  Condens\'ee  (UMR CNRS  7600),  Universit\'e Pierre  et
    Marie  Curie, 4  place  Jussieu,  75252 Paris  Cedex  5 France  \&
    Laboratory J.-V.  Poncelet (UMI CNRS 2615), Independent University
    of  Moscow,  Bolshoy   Vlasyevskiy  Pereulok  11,  119002  Moscow,
    Russia${}^2$}

\begin{abstract}
  A  particle driven  by an  external  force in  a molecular  crowding
  environment  - a  quiescent  bath of  other  particles, makes  their
  spatial distribution inhomogeneous: the bath particles accumulate in
  front of the biased particle (BP) and are depleted behind.  In fact,
  a BP travels together with  the inhomogeneity it creates.  A natural
  question  is  what  will  happen  with  two  BPs  when  they  appear
  sufficiently  close  to each  other  such  that the  inhomogeneities
  around each of them start to  interfere?  In quest for the answer we
  examine here, via  Monte Carlo simulations, the dynamics  of two BPs
  in  a  lattice  gas  of  bath  particles.  We  observe  that  for  a
  sufficiently dense medium, surprisingly,  both BPs spend most of the
  time  together   which  signifies  that  the   interference  of  the
  microstructural  inhomogeneities results  in  effectively attractive
  interactions  between   them.   Such  statistical   pairing  of  BPs
  minimizes  the  size of  the  inhomogeneity  and  hence reduces  the
  frictional  drag force  exerted  on the  BPs  by the  medium.  As  a
  result, in some  configurations the center-of-mass of a  pair of BPs
  propagates faster than a  single isolated BP.  These jamming-induced
  forces  are very different  from fundamental  physical interactions,
  exist  only  in presence  of  an  external  force, and  require  the
  presence of a quiescent bath to mediate the interactions between the
  driven particles.
\end{abstract}

\maketitle

\section{Introduction}

A biased  particle (BP) traveling in  a bath of  particles, which move
randomly  without  any preferential  direction,  drives their  spatial
distribution  out  of  equilibrium.   The bath  particles  accumulate,
creating a  "traffic jam" in front  of the BP and  are depleted behind
it. This  BP can  be, e.g.,  a charge carrier  subject to  an electric
field or  a colloid moved with  an optical tweezer.  The bath particle
may be,  e.g., colloids dispersed  in a solvent or  adatoms performing
activated  hopping  motion  among  the  adsorption sites  on  a  solid
surface.

Such  microstructural changes,  which substantially  enhance  the drag
force  exerted  on  the  BP,  have been  observed  experimentally;  in
particular, in  microrheological measurements of  the drag force  on a
single colloid driven through a $\lambda$-DNA solution \cite{1} or for
a biased  motion of an intruder  dragged into a  monolayer of vibrated
grains  \cite{2}.    Formation  of  an   inhomogeneous  nonequilibrium
distribution has  also been revealed by  Brownian Dynamics simulations
of a  driven colloid in a  $\lambda$-DNA solution \cite{1,3}  and in a
colloidal crystal \cite{4}.   In the latter case, it  was shown that a
large  enough BP  generates a  sufficient stress  to  produce defects,
which  remain localized  near the  BP and  affect the  frictional drag
force.

Microstructural changes of a quiescent medium caused by a biased probe
were  extensively   studied  analytically  \cite{5,6,7,8,9,10,11}  for
hard-core lattice  gases with simple exclusion dynamics,  in which all
particles except  one have symmetric hopping  probabilities, while one
of them - the BP - has a preferential direction of motion.

In one-dimensional systems the size of the jammed region in front of a
BP (as well as  the size of the depleted region in  the wake) grows in
proportion  to  the   traveled  distance.   Thus  the  jamming-induced
contribution to the  frictional drag force $\gamma$ exerted  on the BP
by  the bath  particles  exhibits an  unbounded  growth, $\gamma  \sim
t^{1/2}$,  ($t$  being time),  so  that  the  BP velocity  $V_t^{(1)}$
vanishes\cite{5,6}, $V_t^{(1)} \sim t^{-1/2}$ as $t \to \infty$.  This
insures the validity of  the Einstein relation for anomalous diffusion
in one-dimensional hard-core lattice gases \cite{6,7}.

In higher dimensions, the BP velocity attains a drift value $V^{(1)} =
V^{(1)}_{t  = \infty}$ and  the bath  particle distribution  reaches a
non-equilibrium   stationary  form   \cite{8,9,10,11}.    The  density
profiles are  strongly anisotropic with  a traffic-jam like  region in
front of the BP and a depleted region in its wake.  Strikingly, behind
the BP the bath particle density approaches the mean value $\rho$ as a
power-law of the distance $x$:  $1/x^{3/2}$ and $\ln(x)/x^2$ in 2D and
3D \cite{8,9,10,11},  which signifies that the  medium "remembers" the
passage of  the BP  on large temporal  and spatial scales.   The drift
velocity $V^{(1)}$  and the  jamming-induced drag force  $\gamma$ have
been determined for the  lattice gas model \cite{7,8,9,10,11} and also
for a driven probe in a colloidal mixture \cite{1,12,13}.

The next step  in the understanding of the  jamming-induced forces has
been done in refs.~\cite{14,15,16}.  Dzubiella et al \cite{14} studied
the effective  interactions between  two \textit{fixed} colloids  in a
quiescent viscous solvent exposed to  a flowing bath of small Brownian
particles, while  Krueger and Rauscher  \cite{15} and Khair  and Brady
\cite{16}  considered the  case of  two  colloids \textit{translating}
along  their lines  of centres  with \textit{fixed}  velocities  and a
fixed  distance  apart  of   each  other  in  an  otherwise  quiescent
dispersion  of   noninteracting  colloids.   It   was  realized  that,
remarkably,  microstructural changes  induce effective  forces between
two colloids,  which may be either repulsive  or attractive, depending
on their mutual orientation.

In this  paper we pose a  very natural question within  the context of
microfluidics/microrheology   or  biased   dynamics   under  molecular
crowding  conditions: What  will happen  with two  BPs in  a quiescent
medium of  mutually interacting particles  when both, move not  with a
prescribed  velocity along  some  fixed lines,  but  rather perform  a
biased random motion  subject to some external force?   In contrast to
the  situations  studied in  refs.~\cite{3,14,16},  here  the BPs  can
change their relative position in space and hence, by monitoring their
trajectories we  can understand the overall  effect of microstructural
changes of the medium on the interactions between them.

In order to be as transparent as possible, here we resort to a minimal
model of a hard-core lattice gas of particles whose dynamics obeys the
so-called  simple  exclusion  process  (SEP)  \cite{17,18}.   We  note
parenthetically  that this model  of dynamics  is quite  realistic and
applies to many  physical systems, such as, e.g.,  dynamics of adatoms
on solid  surfaces (see Refs.~\cite{8,9,10,11,17,18}  for more details
and other systems).  We note, as well, that it allows us to single out
the   effect   of   microstructural   changes,   i.e.,   bath-mediated
interactions, and  to exclude possible effects of  solvent (if present)
and solvent-mediated interactions\cite{19} between the BPs.

In this  model the lattice gas  particles - the bath  particles - have
symmetric hopping  probabilities while two  particles - the BPs  - are
subject   to   an  external   force   and   have  asymmetric   hopping
probabilities.    Tracking  the  BPs   trajectories  in   Monte  Carlo
simulations, we observe a  phenomenon of statistical pairing of biased
particles.   We  realize  that  for  sufficiently  dense  systems  the
fraction of time which the second  BP spends at a given point in space
has an apparent  maximum in the vicinity of the  first BP.  Hence, the
interference of non-equilibrium density profiles of the bath particles
formed  around each  of the  BPs  results in  an effective  attractive
interaction between them.  Apart of this, we analyze the properties of
the jamming-induced frictional forces  and determine the velocities of
the BPs appearing in different configurations.

\begin{figure}[!t]
\begin{center}
  \centerline{\includegraphics[width=8.5cm]{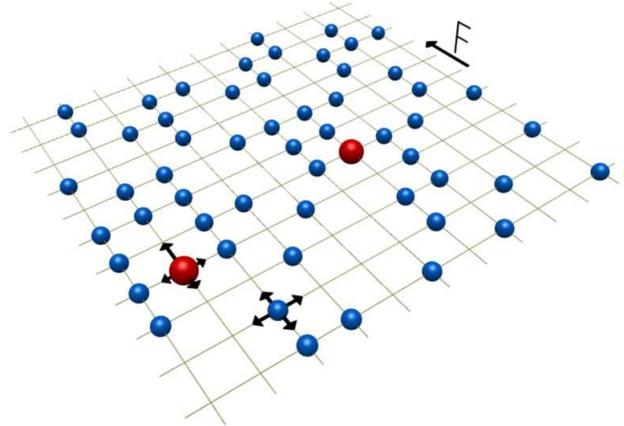}}
  \caption{   \textbf{A  lattice   with   randomly  moving   hard-core
      particles.}   Bath  (blue)   particles  have  symmetric  hopping
    probabilities.  BPs (red) are subject to a constant force and have
    asymmetric hopping probabilities.
\label{fig:model}}
\end{center}
\end{figure}

The paper is outlined as follows: in Section 2 we define the model. In
Section  3 we  focus on  dynamics  of a  single biased  particle in  a
quiescent bath,  describe the density  profiles of the  bath particles
around the BP and determine  the force-velocity relation. In Section 4
we consider  dynamics of two BPs.  Here we define  most probable paths
that  the BPs follow  and describe  the density  profiles of  the bath
particles  forming  around  a  pair  of  BPs  appearing  in  different
configurations. Apart of this,  we define the frictional force exerted
by  the medium on  the BPs  and determine  the velocity  of a  pair of
BPs.  We conclude  in Section  5 with  a brief  recapitulation  of our
results and an outlook of future work.

\section{The Model}

Consider  a square  lattice of  $S \equiv  L_x \times  L_y$  sites, of
spacing $\sigma$ and with periodic boundary conditions. The lattice is
populated  with two  different types  of particles:  $N-M$  {\it bath}
particles and $M$  ($M = 1,2$) BPs (see  figure \ref{fig:model}).  All
particles experience  hard-core interactions, such that  each site can
be either  empty or occupied  by at most  one particle.  The  state of
each site $(X,Y)$ is described by a time-dependent occupation variable
$\eta(X,Y)$;  $\eta(X,Y)=1$  if  the  site  $(X,Y)$  is  occupied  and
$\eta(X,Y)=0 $, otherwise.

Particle dynamics is  defined by the so-called SEP  - simple exclusion
process \cite{17,18}.   Each particle  bears an exponential  clock; in
general, the mean jump-time of the  bath particles and that of the BPs
may be  different; say, it is  $\tau^*$ for the  bath particles, while
for  the BPs  it is  $\tau$.  We  will focus  in what  follows  on the
simplest case  $\tau^* = \tau =  1$ and will only  briefly mention the
effects of different jump times on the friction coefficient.

When  the clock  rings,  a particle  attempts  to jump  from the  site
$(X,Y)$  it occupies  to  one of  the  four nearest-neighboring  sites
$(X',Y')$   according   to  the   normalized   set  of   probabilities
$p(X,Y|X',Y')$.   Once  a  jump   direction  is  chosen,  and  if  the
destination  site is  empty, the  particle moves  to it,  otherwise it
remains on  the site it occupies.  This  stochastic exclusion dynamics
is      a      Markov      process      on     a      state      space
$\mathcal{M}=\{0,1\}^{L_x}\otimes\{0,1\}^{L_y}$.

We  stipulate next that  the dynamics  of the  \textit{bath} particles
obeys a  symmetric SEP (all  hopping probabilities $=1/4$),  while the
BPs are driven by an external  field $\vec{F} = -F \, \hat{e}_{X}$ and
evolve according to an asymmetric SEP:
\begin{equation}
  p({\mathbf{r},\mathbf{r}\pm \sigma \hat{\mathbf{e}}_X}) =
 \mathcal{Z}^{-1}e^{\mp \beta \sigma F/2} \ , \quad
  p({\mathbf{r},\mathbf{r}\pm\sigma\hat{\mathbf{e}}_Y}) =\mathcal{Z}^{-1} \ ,
\end{equation}
where  $\hat{\mathbf{e}}_X = (1,0)$  and $\hat{\mathbf{e}}_Y  = (0,1)$
are  unit  shift vectors,  $\mathbf{r}=(X,Y)$,  $\mathcal{Z}  = 2(1  +
\cosh{(\beta \sigma F/2)})$ and $\beta$ is the inverse temperature.

\section{One biased particle}

To set up  the scene, we focus first  on the case of a single  BP on a
lattice with  $N-1$ bath  particles.  As we  have already  remarked, a
single BP  produces microstructural changes  in the medium  it travels
in,  driving the  spatial distribution  of the  bath particles  out of
equilibrium.  To  quantify the microstructural changes  of the medium,
we   consider  the  following   realization-dependent  "inhomogeneity"
measure:
\begin{equation} \label{eq:gf}
\g_0(\mathbf{r};t) = \frac{1}{t}\int_0^td\tau
\left(\frac{1}{\rho}\sum_{i=2}^{N} \delta\left(\mathbf{r}_i(\tau) -
  \mathbf{R}_1(\tau)-\mathbf{r}\right)\right) \ ,
\end{equation}
where $\delta(\cdot)$  is the  Kronecker-delta, which equals  $1$ when
its argument is $0$ and  is zero otherwise, $\rho= (N-1)/(S-1)$ is the
mean  density  of   bath  particles,  while  $\mathbf{R}_1(\tau)$  and
$\mathbf{r}_i(\tau)$ denote the positions of  the BP and of the $i$-th
bath  particle  at time  $\tau$,  respectively,  for  a given  set  of
realizations    of     their    trajectories.     An     average    of
$\g_0(\mathbf{r};t)$   over  different   realizations   of  particles'
trajectories defines the time-averaged van Hove function\cite{20}.

The    realization-dependent   functional    $\g_0(\mathbf{r};t)$   in
Eq.\ref{eq:gf}  defines the fraction  of time  the site  $\mathbf{r} =
(x,y)$, in the  frame of reference moving with the  BP, is occupied by
bath  particles  during  a  time   $t$  for  a  given  realization  of
trajectories  of  the bath  particles  and  the BP.\footnote{Here  and
  henceforth, small  characters $x$ and  $y$ will denote  a coordinate
  system  moving  with the  BP,  while $X$  and  $Y$  will denote  the
  laboratory frame of reference.}   If the spatial distribution of the
bath  particles  converges  to  a  stationary  form,  {\it  i.e.},  if
$\lim_{t\to\infty} \g_0(\mathbf{r};t) = \g_0(\mathbf{r})$ exists, then
$\rho \g_0(\mathbf{r})$ can  also be thought of as  the bath particles
density  profile   as  seen  from  stationary   moving  BP.   Clearly,
$\g_0(\mathbf{0})=0$    due   to    the   hard-core    exclusion   and
$\g_0(\mathbf{r})  \to 1$  when  $|\mathbf{r}| \to  \infty$.  We  have
conveniently  normalized $\g_0(\mathbf{r};t)$ to  $\rho$, so  that any
deviation  $\g_0\ne1$  indicates  a  non  zero  dynamical  correlation
between the BP and the medium.

\begin{figure}[!t]
  \centering
  \includegraphics[width=8.5cm]{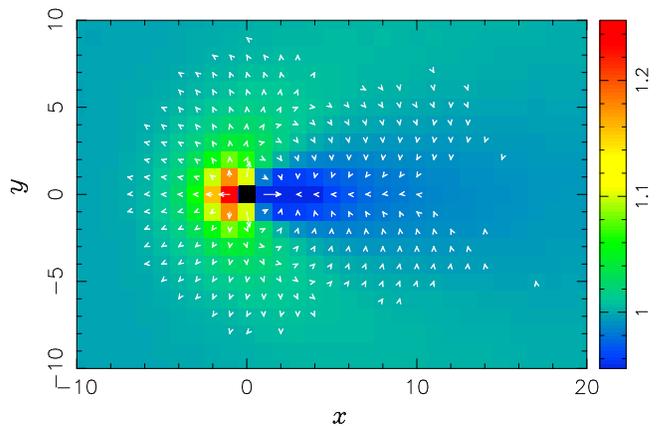}
  \caption{\textbf{Microstructural changes of the medium produced by a
      single BP.}  The profile  $\g_0(x,y)$ in Eq.\ref{eq:gf} is shown
    for a lattice comprising  $61\times21$ sites at density $\rho=1/2$
    and  $\beta  \sigma F=5$.   The  vector  field  shows the  average
    velocity  field of  the bath  particles defined  in  the reference
    frame of the BP (a black square).  Velocity vectors with magnitude
    less than $0.0005$ are not plotted.
\label{fig:gmed}}
\end{figure}

In  figure  \ref{fig:gmed}  we   depict  the  bath  particles  density
$\g_0(\mathbf{r})$ using a colour  map.  The density profiles around a
stationary  moving  BP are  characterized  by  a jammed,  high-density
region in front of the BP and a pronounced region depleted by the bath
particles past  the BP.  This  agrees quite well with  the theoretical
prediction   of  Refs.\cite{7,8,9,10}.   We   verify,  as   well,  the
theoretical  prediction that the  density past  the BP  approaches the
average value $\rho$ not exponentially with the distance $x$, but as a
slow  power-law  $x^{-3/2}$.  Moreover,  in  figure \ref{fig:gmed}  we
superimpose  the  average velocity  field,  which  shows  that the  BP
induces a  regular global motion  of the bath  particles predominantly
towards  the regions  with lower  density.  Note  that  similar motion
patterns  have  been  observed  experimentally for  biased  motion  in
granular media \cite{2}.

\begin{figure}[!t]
  \centering
\includegraphics[width=7.5cm]{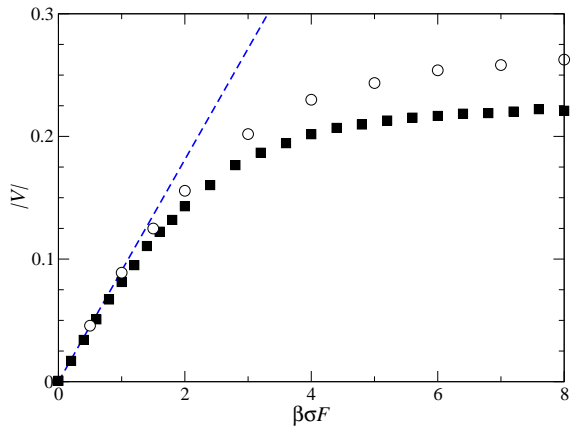}
\caption{\textbf{Force-velocity relation.} Drift velocity $V^{(1)}$ of
  a  single  BP  vs $F$  (solid  squares).   The  dashed line  is  the
  theoretical  prediction  $V^{(1)} =  F/\xi$  in  which the  friction
  coefficient  $\xi$  is  given by  Eq.\ref{eq:force-velocity}.   Open
  circles denote  the results of  the Monte Carlo simulations  for the
  drift velocity  of a pair of  BPs in the  $(2,0)$ configuration (see
  the explanations in Section 4.2). \label{fig:force-velocity}}
\end{figure}

Next, in  figure \ref{fig:force-velocity} we depict  the dependence of
the  BP's  drift  velocity  $V^{(1)}$  on the  applied  force  $F$,
i.e. the force-velocity relation, for $\rho = 1/2$.

The notable  feature of the  observed force-velocity curve is  that in
the   limit   of  sufficiently   small   forcing  (sufficiently   high
temperatures)  the   drift  velocity  shows   a  Stokesian,  linear
dependence on  the applied force,  $V^{(1)} = F/\xi$,  which signifies
that in this limit the  frictional force exerted by the bath particles
on the BP is viscous.

According   to   Refs.\cite{7,8,9,10}   (see   also  Section   VI   in
Ref.\cite{21}), in  this linear regime the  friction coefficient $\xi$
can be expressed as a sum of two contributions,
\begin{equation}
\xi = \xi_{mf} + \xi_{coop},
\label{eq:force-velocity}
\end{equation}
where the first term,
\begin{equation}
\xi_{mf} = \frac{4 \tau}{\beta \sigma^2 (1 - \rho)},
\end{equation}
is  essentially  a  mean-field  result corresponding  to  a  perfectly
stirred  monolayer; one  may interpret  $(1- \rho)/\tau$  just  as the
frequency of  the BPs "successful" jump  events. The second  term is a
"jamming-induced" contribution:
\begin{equation}
\xi_{coop} = \frac{4 \tau^*}{\beta \sigma^2 (1 - \rho)} \,
\frac{(\pi - 2) \rho}{1+  (1 - \rho) \tau^*/\tau}
\end{equation}
stemming out of a cooperative behavior in the monolayer - a non-linear
interplay between the BP dynamics and the formation of non-equilibrium
density  profiles around  it. Analogous  results for  $\xi$  have been
obtained for  three-dimensional\cite{11} and one-dimensional\cite{5,6}
systems; in the latter case $\xi$ diverges as $t \to \infty$.

Dividing $\xi_{mf}$ by $\xi_{coop}$, we have
\begin{equation}
\frac{\xi_{mf}}{\xi_{coop}} \sim \frac{1 - \rho}{\rho} + \frac{1}{\rho}
\frac{\tau\phantom{^*}}{\tau^*}.
\end{equation}
One  notices that  the  jamming-induced contribution  to the  friction
coefficient dominates  when $\tau^* \gg  \tau$ and the  bath particles
mean density is not too small. Conversely, the mean-field contribution
is  clearly  the  dominant  one  when  $\rho \ll  1$  or  $\tau^*  \ll
\tau$.  For  moderate   densities,  $\xi_{mf}$  and  $\xi_{coop}$  are
comparable.

\begin{figure}[!b]
  \centering
  \includegraphics[width=8cm]{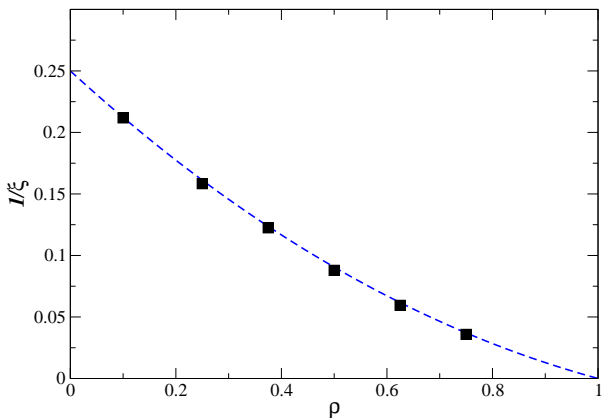}
  \caption{\textbf{Friction  coefficient  $\xi$  versus density.}  The
    dashed     line    is     the     theoretical    prediction     in
    Eq.\ref{eq:force-velocity}, while the symbols define the numerical
    simulations results for $\sigma=\beta=1$ and $\tau=\tau^*=1$.
\label{fig:mobility-vs-density}}
\end{figure}

Finally, in figure \ref{fig:mobility-vs-density} we compare the result
in Eq.\ref{eq:force-velocity}  against Monte Carlo  simulation results
for the  slope of  the force-velocity relation  in the limit  of small
forcing, at different  densities $\rho$ and $\tau =  \tau^* = 1$.  One
observes a  very good agreement  between a theoretical  prediction and
numerical data.

\section{Two biased particles}

We turn  now to the  case of primary  interest - two BPs.   We suppose
that initially the BPs are  placed at sites $(X,Y)$ and $(X',Y')$, and
that $X < X'$; following the terminology of Khair and Brady \cite{16},
we then  refer to the particle  initially at $(X,Y)$  as the "leading"
BP,  and  the  one  at  $(X',Y')$  - the  "trailing"  BP  (see  figure
\ref{fig:model}). We recall that  the coordinate system defined in the
reference frame of the leading BP is be denoted by $(x,y)$.

\subsection{Mutual orientation of the BPs}
We concentrate first on the analysis of the most probable paths of the
trailing  BP in  the  reference frame  of  the leading  BP.  For  this
purpose, we study numerically the behaviour of a realization-dependent
functional
\begin{equation}\label{eq:ge}
\g(\mathbf{r};t) = \frac{1}{t - t_0}\int_{t_0}^td\tau
\delta\left(\mathbf{R}_2(\tau)-\mathbf{R}_1(\tau)-\mathbf{r}\right) \ ,
\end{equation}
where  $\mathbf{R}_1(\tau)$  and  $\mathbf{R}_2(\tau)$ stand  for  the
instantaneous  positions  of   the  leading  and  trailing  particles,
respectively,   for  a  given   realization  of   their  trajectories.
Similarly    to   the    functional    defined   in    Eq.\ref{eq:gf},
$\g(\mathbf{r};t)$ defines, for a given realization of the leading and
trailing  BPs  trajectories, the  fraction  of  time  during the  time
interval  $t -  t_0$  that the  site  $\mathbf{r}$, (in  the frame  of
reference  moving with  the  leading  BP), has  been  occupied by  the
\textit{trailing} BP.

\begin{figure*}[t!]
  \centering
  \includegraphics[width=13.5cm]{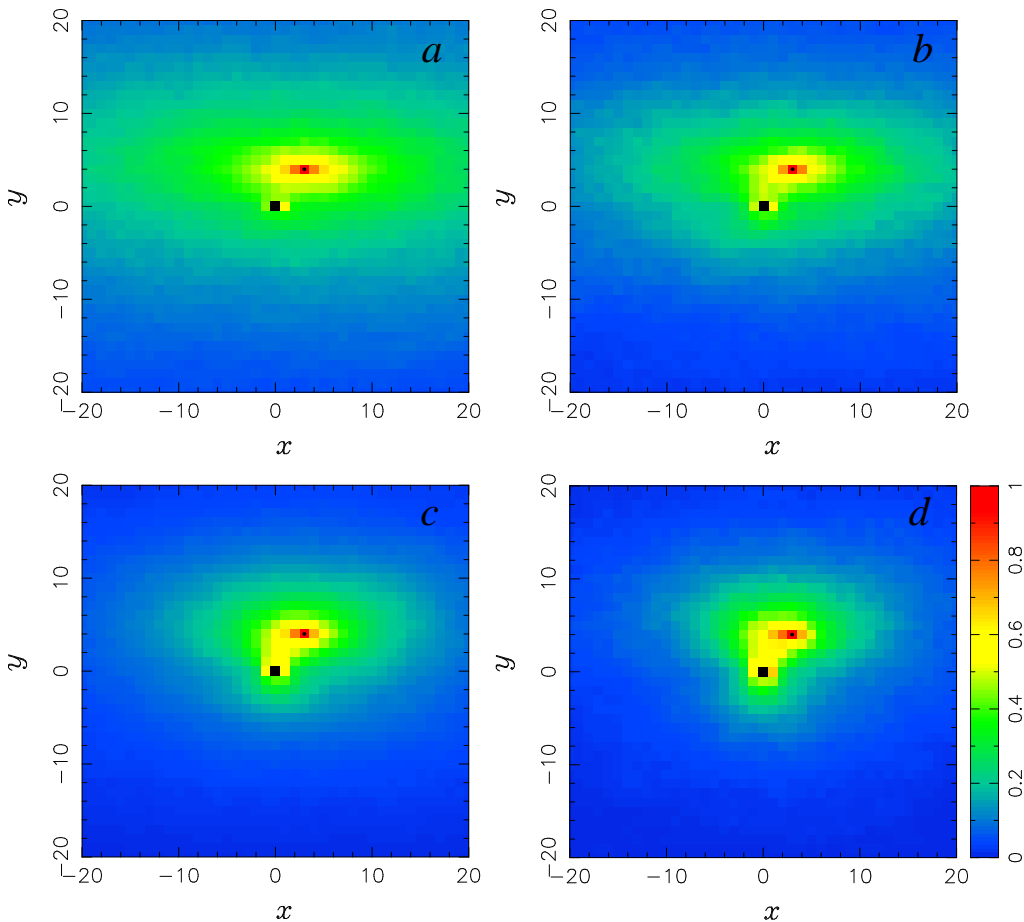}
  \caption{\textbf{Most probable paths.}  The profile of average local
    occupation times $<\!\!\g(\mathbf{r};t)\!\!>$, Eq.\ref{eq:ge}, for
    the  trailing  BP  commencing at  $\mathbf{R}_2(0)=(3,4)$  (little
    black knot),  $\beta\sigma F=5$ and densities  $\rho$: $a$) $1/4$,
    $b$)  $1/2$, $c$)  $3/4$ and  $d$) $0.9$.   High values  of $<\!\!
    \g(\mathbf{r};t)\!\!>$ (yellow)  indicate the most  probable paths
    that the trailing BP follows.
    \label{fig:paths}}
\end{figure*}

In simulations the  leading BP is initially placed  at the origin, the
trailing one  is placed at position $\mathbf{R}_2(0)$,  while the bath
particles are placed at random,  with mean density $\rho$, on the rest
of the lattice.  We let the system evolve (the bath particles follow a
standard symmetric SEP while the BPs dynamics obeys an asymmetric SEP)
for $10^8$ time steps, until the density profile of the bath particles
around  two  BPs attains  a  stationary  form.   After this  transient
period, we  define the  moment $t_0$ when  the trailing  BP re-appears
again at site $\mathbf{R}_2(0)$.  Then, during the next $t-t_0 = 10^6$
time steps  we evaluate $\g(\mathbf{r};t)$ by recording  the number of
times each site $\mathbf{r}$ in  the frame of reference of the leading
BP has been visited by the  trailing BP within this realization of the
process.

In figure  \ref{fig:paths} we plot the average  local occupation times
$<\!\!   \g(\mathbf{r};t)\!\!>$\footnote{The   angle  brackets  denote
  averaging over  different realizations  of the leading  and trailing
  BPs trajectories.  We consider $10^3$ such realizations.} for $\beta
\sigma F = 5$ and $\mathbf{R}_2(0)=(3,4)$.  Our results show that when
the inhomogeneities  around each  BP do not  overlap, both  BPs travel
almost  independently.   In this  case,  the  profile  of the  average
occupation times  around the  initial position of  the trailing  BP is
almost  symmetric (see Fig.~\ref{fig:paths}:$a$)  with a  small second
maximum just  after the leading  BP.  The overlap  becomes significant
for  either sufficiently  high  density $\rho$,  larger driving  force
$\beta \sigma F$  or naturally, when the leading  and the trailing BPs
are close  enough.  For progressively  higher densities of  the medium
particles (see Figs.\ref{fig:paths}:$b$-$d$) we observe a considerable
qualitative     change    in     the    form     of     the    profile
$<\!\!\g(\mathbf{r};t)\!\!>$: it  becomes considerably more asymmetric
and is  characterized by an  apparent "bridge" connecting  the leading
and the trailing BPs.  Hence,  the probability of finding the trailing
BP in the vicinity of the leading one is getting progressively higher.

\begin{figure}[t!]
  \centering
  \includegraphics[width=8.5cm]{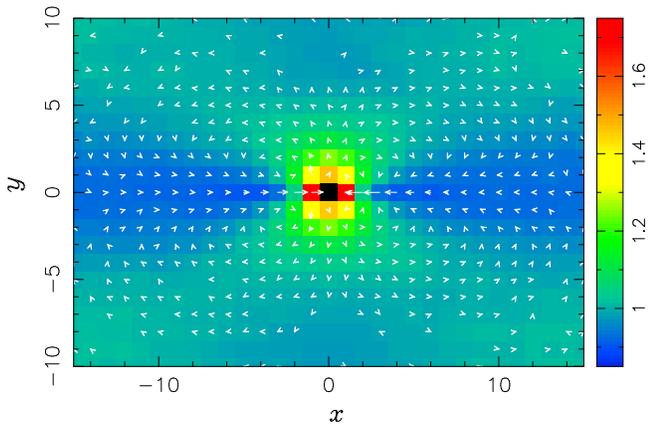}
  \caption{\textbf{Local occupation times  of the trailing BP.} Colour
    map  of the  occupation time  of  the trailing  BP given  by
    $\g(\mathbf{r})/(S-1)$,  for a lattice  comprising $31  \times 21$
    sites,  $\rho =  1/2$ and  $\beta \sigma  F=5$.   The superimposed
    vector  field  superimposed shows  the  average relative  velocity
    field of  the trailing BP measured  in the reference  frame of the
    leading BP.  Velocity vectors with magnitude less than $0.001$ are
    not plotted.
    \label{fig:gbias}}
\end{figure}

Further on, we focus on  a \textit{single} (very long) trajectory of a
trailing BP (see Fig.~\ref{fig:gbias}).   We let the system evolve for
$10^8$  time steps  to  ensure that  the  bath particles  distribution
around  the two  BPs reaches  a  stationary form.   Then, we  evaluate
$\g(\mathbf{r};t)$ by  tracking the trajectory  of the trailing  BP in
the frame  of reference of the  leading one. We have  checked that for
sufficiently    large   times    (in   our    simulations   $t=10^8$),
$\g(\mathbf{r};t)$     converges    to    a     stationary    function
$\g(\mathbf{r})$, which moreover, is independent of the initial state.

Figure    \ref{fig:gbias}   shows    the   local    occupation   times
$\g(\mathbf{r};t)$ of  a single trajectory together  with the relative
velocity field  of the  trailing BP with  respect to the  leading one.
The circulating  structure of  this field indicates  that the  pair is
statistically stable,  presenting a stable  direction ($x$-axis) along
which  the  trailing and  the  leading  BPs  "pair", and  an  unstable
direction ($y$-axis) along  which the trailing BP moves  away from the
leader.

This signifies  that when the  density profiles of the  bath particles
emerging   around  each   of   the  BPs   start   to  interfere,   the
jamming-induced  frictional force  exerted on  the trailing  BP  is no
longer parallel to the $x$-axis  (direction of the external force $F$)
but is tilted  by some angle pointing towards the  wake of the leading
BP.   As  a consequence,  the  trailing  BP  experiences an  effective
attraction  towards  the  leading   one  such  that  both  driven  BPs
statistically  "pair". From the  microscopic dynamics  viewpoint, this
interference modifies locally the statistics of successful hops of the
trailing BP, increasing the likelihood of hops towards the wake of the
leading BP.

\begin{figure}[b!]
  \centering
  \includegraphics[width=8cm]{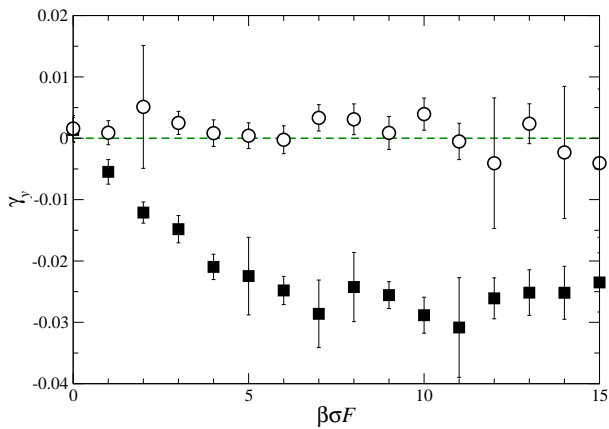}
  \caption{\textbf{Force exerted  on the trailing  BP.}  Time averaged
    jamming-induced force  experienced by the trailing  BP at position
    $(0,7)$  (open  circles)  and  $(4,2)$  (solid  squares),  in  the
    reference frame of leading one, as a function of $\beta \sigma F$.
    The error bars indicate the standard deviation.
    \label{fig:force}}
\end{figure}

To  substantiate   this  claim,   we  have  computed   numerically  an
instantaneous        jamming-induced        nonequilibrium       force
$\vec{\gamma}(\mathbf{r};t)$ that we define as
\begin{equation} \label{gamma}
\vec{\gamma}(\mathbf{r};t) = \sum_{i=\{x,y\}}
\left(\eta(\mathbf{r(t)}-\hat{e}_i) -
  \eta(\mathbf{r(t)}+\hat{e}_i)\right)\hat{e}_i ,
\end{equation}
where $\eta(\mathbf{r(t)})$ are the local occupation variables defined
in Section II. Clearly, $\vec{\gamma}(\mathbf{r};t)$ is the force that
is felt by the trailing BP  being at site $\mathbf{r}$ with respect to
the leading BP at time moment $t$.

On  average, $\vec{\gamma}(\mathbf{r})$  will be  different  from zero
only if the  distribution of the bath particles  is inhomogeneous.  In
general,      as     follows     from      figure     \ref{fig:gbias},
$\vec{\gamma}(\mathbf{r})$  should  depend  on  the  position  of  the
trailing   BP  with   respect   to  the   leading   one.   In   figure
\ref{fig:force} we show  the time averaged $y$ component  of the force
$\vec{\gamma}$ that the trailing BP experiences when it is at position
$(0,7)$ (circles)  and $(4,2)$ (squares)  with respect to  the leader.
In the  first case, the trailing BP  is far enough from  the leader so
that the net force along  $y$ is numerically zero, irrespective of the
strength  of the  field.  On  the contrary,  in the  second  case, the
trailing BP  is in  the stable basin  of attraction and  $\gamma_y$ is
negative,  indicating that  (in agreement  with Fig.~\ref{fig:gbias}),
the force pushes  the trailing BP towards the  leading one.  For small
fields $\gamma_y$  grows linearly with $\beta \sigma  F$ and saturates
at   larger  fields.    Of  course,   the  amplitude   of   the  force
$\vec{\gamma}(\mathbf{r})$ depends  on the density  of bath particles,
external force $F$ and,  naturally, on the temperature, which controls
the rate  of the bath  particles' migration and thus  their capability
for smoothing down the inhomogeneities created by the BPs.

\subsection{The BPs velocities}
The (mean) drift velocity $V^{(1)}$ of a single isolated BP is totally
determined  by $\beta  \sigma  F$, $\rho$  and  the rate  of the  bath
particles'  migration\cite{9,10}.   For  two  BPs,  when  they  appear
sufficiently close to each  other such that the inhomogeneities around
each  of them  start  to  interfere, their  drift  velocities and  the
velocity  of their center  of mass  will also  depend on  their mutual
orientation. One may  expect that only at large  mutual separations of
the BPs their velocities are equal to $V^{(1)}$.

\begin{figure*}[t!]
  \centering
  \includegraphics[width=17cm]{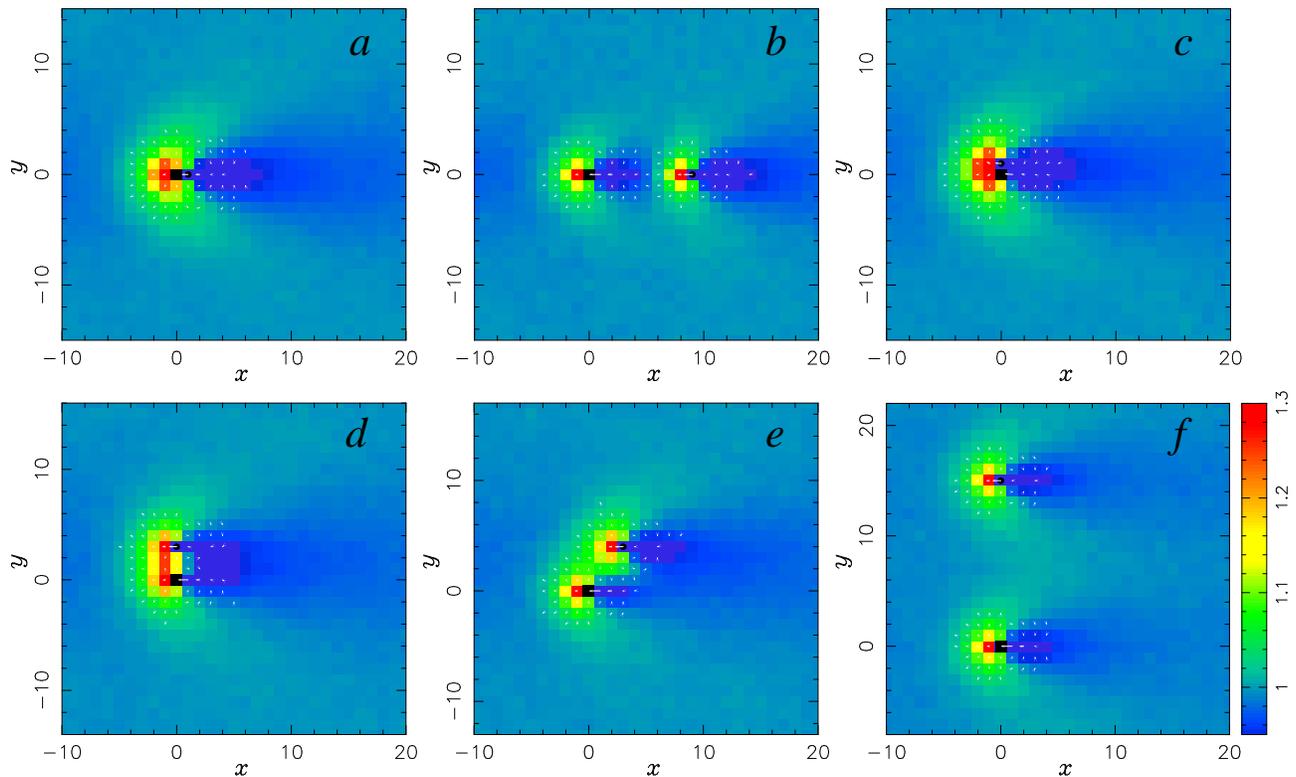}
  \caption{\textbf{Microstructural  changes   produces  by  two  BPs.}
    Colour  map  of  the  mean  local occupation  times  of  the  bath
    particles    $\langle   g_0(x,y)\rangle$,   for    six   different
    configurations of two  BPs, defined in the reference  frame of the
    leading  BP (black  square), with  $\rho =  1/2$  and $\beta\sigma
    F=5$.  The  arrows define  the vector velocity  field.  Velocities
    with magnitude less than $0.005$ are not plotted.\vspace{4mm}
    \label{fig:inter}}
\end{figure*}

To clarify this  issue, we have studied the  velocities of the leading
and trailing BPs at a fixed force  $\beta \sigma F = 5$ and at a fixed
bath-particle  density   $\rho  =  1/2$,  but   for  different  mutual
orientations.  Six different mutual  orientations of the BPs, together
with  the corresponding  microstructural  changes of  the medium,  are
presented in figure \ref{fig:inter}. In this figure we also depict the
velocity field of the bath particles.

\begin{figure*}[t!]
  \centering
  \includegraphics[width=17cm]{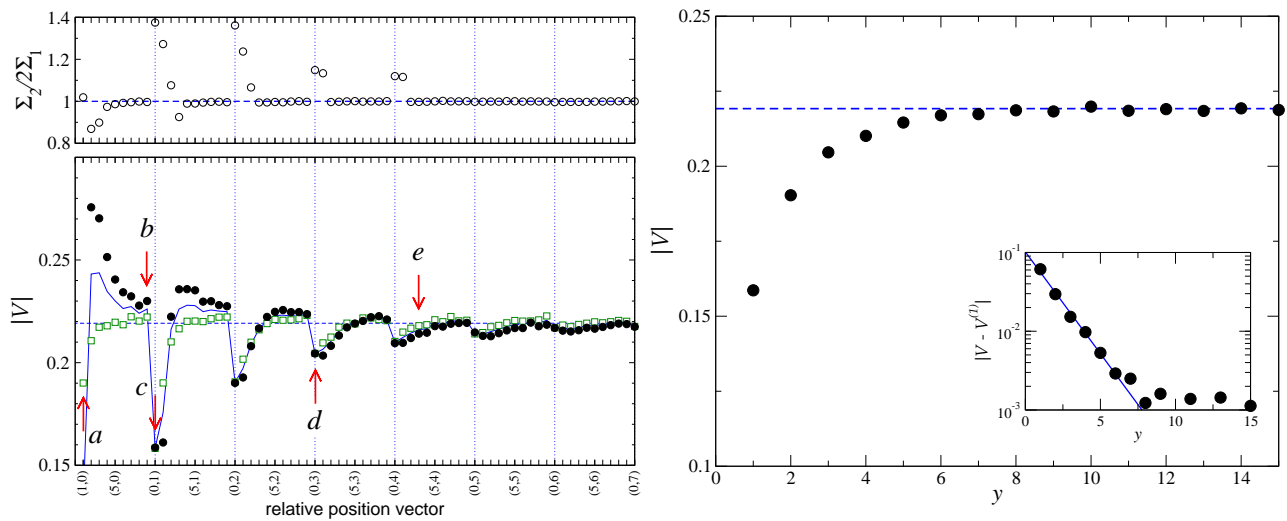}
  \caption{\textbf{Drift   velocities   of   the  BPs   in   different
      configuration} Left lower panel.- Magnitude of the $x$-component
    of the velocity of the  leading (open squares) and trailing (solid
    circles) BPs for different  mutual orientations, $\beta \sigma F =
    5$ and  $\rho = 1/2$.  The  blue dashed line  defines the velocity
    $V^{(1)}$  of a  single isolated  BP,  while the  solid blue  line
    defines the center-of-mass velocity of  a pair of BPs.  The arrows
    indicate    some    of     the    configurations    depicted    in
    figure~\ref{fig:inter}.   Left   upper  panel.  -   The  ratio  of
    "effective"  inhomogeneities created by  two BPs  and a  single BP
    (see the text for the explanations).  Right panel.- Drift velocity
    of the transversal configurations  as a function of the separation
    $y$.   The  inset  shows  that  the drift  velocity  converges  to
    $V^{(1)}$ exponentially.
    \label{fig:vel}}
\end{figure*}

Further on,  in figure ~\ref{fig:vel} we plot  the instantaneous drift
velocities of  the leading and trailing  BPs as the  function of their
mutual  orientation. In the  left panels,  along the  horizontal line,
each "tick" corresponds to the  position $(x,y)$ of the trailing BP in
the reference frame of the  leading one. Between each pair of vertical
dotted lines, we place nine points with fixed $y$ and $x$ varying from
$1$ to  $9$, i.e., $(1,y)$, $(2,y)$  to $(9,y)$.  From  the left lower
panel one indeed concludes that at large separations the velocities of
the BPs are nearly the  same, and coincide with the velocity $V^{(1)}$
(blue dashed line) of a  single isolated BP.  Conversely, when the BPs
are close  enough, their  velocities may be  very different  from each
other, as well as from $V^{(1)}$.

At  a  fixed  $y$,  the  velocity  of the  trailing  BP  is  always  a
non-monotonic function of its $x$-coordinate: it is always minimal for
$x  = 0$,  grows  abruptly  with the  $x$-coordinate  and then  decays
towards $V^{(1)}$.  The leading BP velocity, at a fixed $y$-coordinate
of the trailing BP, is a growing function of the $x$-coordinate of the
latter.  Despite the  fact that the configuration $(1,0)$  is the most
probable (see figure ~\ref{fig:gbias}),  the velocity of the center of
mass (solid blue  line in figure~\ref{fig:vel}) of such  a pair is not
the largest  one, which is  a bit counter  intuitive.  As a  matter of
fact, this is the consequence of the hard-core interaction between the
leading and trailing BPs that hinder the motion of the latter reducing
the average  velocity of the pair.   The blocking effect  is no longer
present in the next configuration $(2,0)$ for which (together with the
configuration  $(3,0)$), the  velocity  of the  center-of-mass is  the
largest.   This signifies  that such  a pair  of the  BPs  creates the
smallest  microstructural  changes, and  in  response, encounters  the
least  jamming-induced  frictional  force,  hence the  least  possible
dissipation.

Curiously enough, the velocity of the center-of-mass of such a pair is
always higher than  the velocity of a single isolated  BP. This can be
seen in figure \ref{fig:force-velocity}, in which we compare numerical
force-velocity relations for a single BP (solid squares) and a pair in
the configuration $(0,2)$ (open  circles), for different values of the
driving force.   The center-of-mass velocity  is also higher  than the
velocity  of a  single  isolated BP  for  the configurations  $(3,0)$,
$(4,0)$ to $(9,0)$, $(2,1)$ to $(9,1)$ and $(4,2)$ to $(9,2)$.  On the
contrary, the microstructural changes of the medium induce the largest
frictional  force  on  pairs  $(0,y)$,   for  which  the  BPs  are  in
perpendicular orientation  with respect  to the field  direction, such
as,  e.g.,  configurations  (c)  and (d).   These  configurations  are
indicated  in figure~\ref{fig:vel}  by  vertical arrows.   Due to  the
symmetry of the problem, in these configurations the velocities of the
leading and trailing  BPs coincide, and are smaller  than the velocity
of a single isolated BP.

To better understand  such a behaviour we study a  measure of the size
of the inhomogeneity created by the BPs.  This configuration-dependent
measure that we  call $\Sigma$, equals the sum  of the mean occupation
times  of the  bath particles,  over the  lattice sites  on  which the
absolute deviation  of $\langle  g_0(x,y)\rangle$ from the  mean value
$\rho$ exceeds a certain threshold $\mathcal{G}$:
\begin{equation}
\Sigma = \sum_{ \delta g > \mathcal{G}} \langle g_0(x,y)\rangle\ ,
\label{eq:Sigma}
\end{equation}
where $\delta  g = |\langle g_0(x,y)\rangle  - \rho|$. To  look at the
correlation   between  the   size   of  the   inhomogeneity  and   the
jamming-induced interaction among  different BPs, we denote $\Sigma_n$
($n=1,2$), as the  inhomogeneity measure \eref{eq:Sigma}, computed for
$n$ BPs. In the left top panel of figure ~\ref{fig:vel} we present the
ratio  $\Sigma_2/2\Sigma_1$ for different  configurations of  the BPs,
clearly showing  that the largest  (smallest) velocities of a  pair of
BPs occur  for smallest (largest) values  of $\Sigma_2/2\Sigma_1$.  In
particular, for the  pair configurations whose center-of-mass velocity
is larger  than the  single BP, we  obtain $\Sigma_2/2 \Sigma_1  < 1$,
indicating that the overall size  of the inhomogeneity produced by the
pair of BPs  in such configurations is smaller  than the corresponding
size of two isolated BPs.

Finally, in  the right panel  of figure~\ref{fig:vel} we show  that at
large  separations  along  the  $y$-axis  the  velocity  of  the  pair
approaches the velocity  of a single isolated BP  exponentially, as is
the  case  of  configuration  (f), corresponding  to  the  orientation
$(0,15)$.

\section{Conclusions}

To   recap,  we   have  addressed   here  a   problem   of  effective,
non-equilibrium interactions that emerge  between two driven probes in
a medium  of randomly moving hard-core  (but otherwise noninteracting)
particles - a  quiescent bath.  We have shown  that for a sufficiently
dense medium  the probes experience an {\it  attractive} force towards
each other, leading to their statistical pairing.  Such a pairing sets
in when the non-equilibrium inhomogeneities in the distribution of the
bath  particles,  created by  the  probes,  start  to interfere.   The
inhomogeneities around each driven  probe decay exponentially with the
distance from the probe, except for the wake of the probe in which the
decay is algebraic.   Consequently, these non-equilibrium interactions
are anisotropic and typically  short-ranged, except for the situations
when the second driven probe appears in the wake of the first one.

The formation of pairs reduces  the overall size of the inhomogeneity,
minimizing the frictional drag force  the medium exerts on each probe.
As  a result,  in some  configurations  the center-of-mass  of a  pair
propagates  faster than  a  single isolated  BP.  The  jamming-induced
forces, which emerge in the  situation under study, are very different
from fundamental  physical interactions, exist only in  presence of an
external  force, and  require  the  presence of  a  quiescent bath  to
mediate the interactions between the driven intruders.

We note that our results  have been obtained for a somewhat simplified
model of a non-interacting lattice gas with simple exclusion dynamics,
which  allowed us  to single  out  the effect  of the  jamming-induced
interactions.    This   model    can   be   generalized   in   several
directions. First of all, one  may consider a situation appropriate to
a colloidal  solution, when some  solvent is present.   Solvent itself
will produce  long-range hydrodynamic interactions  between the driven
probes and correlate dynamics of the bath particles (see Ref.\cite{19}
and references therein).  One may  expect that the pairing effect will
become more  pronounced in this  case. Second, we have  considered the
case of just  two driven probes. It might be  interesting to study the
specific features of  pairing in the situation when  there are many of
them -  the effect observed here  seems very much  like "an elementary
act"  for  the  phenomenon  of  lane  formation  in  partially  driven
colloids\cite{22,23}.   Finally, we  note  that we  have focused  here
solely on the case when the  biased motion of the BPs results from the
presence of  an external force acting  on them.  For  biased motion in
intracellular media  or under molecular crowding  conditions, it might
also be  interesting to consider  other types of biased  motion, e.g.,
the cases  of self-propelled  particles or swimmers.   These important
situations merit further investigation.

\section*{Acknowledgments}
We  thank  J.  Talbot,  J.   L.  Lebowitz  and  H.  Spohn  for  helpful
discussions.   C.   M.-M.   acknowledges  support  from  the  European
Research  Council and  the  Academy of  Finland.   G.O.  is  partially
supported  by  Agence Nationale  de  la  Recherche  (ANR) under  grant
``DYOPTRI''.

\bibliography{paper} 

\end{document}